\documentclass[12pt,preprint]{aastex} 
\shorttitle{Hot Subdwarf Companion of KOI-81} 
\shortauthors{Matson et al.} 
 
\begin{document} 
 
\received{2015 March 9} 
\accepted{2015 May 4} 
 
\title{HST/COS Detection of the Spectrum of the Subdwarf Companion 
of KOI-81\footnote{Based on observations made with the NASA/ESA Hubble Space Telescope, 
obtained at the Space Telescope Science Institute, which is operated by the 
Association of Universities for Research in Astronomy, Inc., under NASA contract 
NAS 5-26555. These observations are associated with program \#12288.}} 
 
\author{Rachel A. Matson\altaffilmark{2}, Douglas R. Gies, Zhao Guo, Samuel N. Quinn} 
\affil{Center for High Angular Resolution Astronomy and  
 Department of Physics and Astronomy,\\ 
 Georgia State University, P. O. Box 5060, Atlanta, GA 30302-5060, USA; \\ 
 rmatson@chara.gsu.edu, gies@chara.gsu.edu, guo@chara.gsu.edu, quinn@astro.gsu.edu} 
 
\author{Lars A. Buchhave\altaffilmark{3}, David W. Latham}
\affil{Harvard-Smithsonian Center for Astrophysics, Cambridge, MA 02138, USA; \\
 lbuchhave@cfa.harvard.edu, dlatham@cfa.harvard.edu}

\author{Steve B. Howell\altaffilmark{2} and Jason F. Rowe\altaffilmark{4}}
\affil{NASA Ames Research Center, P. O. Box 1, M/S 244-30, Moffett Field, CA 94035, USA; \\
 steve.b.howell@nasa.gov, Jason.Rowe@nasa.gov}

\altaffiltext{2}{Visiting Astronomer, Kitt Peak National Observatory,
 National Optical Astronomy Observatory, operated by the Association
 of Universities for Research in Astronomy, Inc., under contract with
 the National Science Foundation.}

\altaffiltext{3}{Centre for Star and Planet Formation, Natural History Museum of Denmark, 
 University of Copenhagen, DK-1350 Copenhagen, Denmark}

\altaffiltext{4}{SETI Institute, Mountain View, CA 94043, USA}

\slugcomment{Accepted for ApJ; 05/04/2015} 

 
\begin{abstract} 
KOI-81 is a totally eclipsing binary discovered by the {\it Kepler} mission
that consists of a rapidly rotating B-type star and a small, hot companion. 
The system was forged through large scale mass transfer that stripped the 
mass donor of its envelope and spun up the mass gainer star.  We present 
an analysis of UV spectra of KOI-81 that were obtained with the Cosmic Origins
Spectrograph on the {\it Hubble Space Telescope} that reveal for the 
first time the spectral features of the faint, hot companion.  
We present a double-lined spectroscopic orbit for the system that 
yields mass estimates of $2.92 M_\odot$ and $0.19 M_\odot$ for the 
B-star and hot subdwarf, respectively.  We used a Doppler tomography
algorithm to reconstruct the UV spectra of the components, and a
comparison of the reconstructed and model spectra yields effective temperatures 
of 12 kK and 19 -- 27 kK for the B-star and hot companion, respectively.
The B-star is pulsating, and we identified a number of peaks in the 
Fourier transform of the light curve, including one that may indicate 
an equatorial rotation period of 11.5~hours.  The B-star has an 
equatorial velocity that is $74\%$ of the critical velocity where
centrifugal and gravitational accelerations balance at the equator, and 
we fit the transit light curve by calculating a rotationally distorted 
model for the photosphere of the B-star.
\end{abstract} 
 
 
 
\setcounter{footnote}{0} 
 
\section{Introduction}                              
 
The NASA {\it Kepler} mission of precise and long time span photometry
has led to many surprising discoveries.  Among the initial set 
of remarkable targets, Rowe et al.\ (2010) reported on observations
of two transiting systems dubbed KOI-74 ($P=5.2$~d) and 
KOI-81 ($P=23.9$~d) that display light curves with minima that 
were deeper during occultation than during transit, implying that the 
planetary size companions are hotter than their A- or B-type host stars.  
Rowe et al.\ (2010) proposed that these companions were actually 
very low mass white dwarf (WD) stars, the remnants of more massive 
progenitors in an interacting binary.  The light curves of both 
systems were further analyzed by van Kerkwijk et al.\ (2010) 
who found that both appear more luminous during orbital 
phases of the bright star approach due to relativistic Doppler boosting
(``beaming binaries''; Zucker et al.\ 2007).  Van Kerkwijk et al.\
used the radial velocity semiamplitude implied by Doppler boosting
to estimate that the white dwarf stars had masses of $\approx 0.2 - 0.3 M_\odot$. 

The {\it Kepler} observations eventually led to the discovery of 
other similar eclipsing binary systems.  Carter et al.\ (2011) found 
that KIC~10657664 consists of a hot WD orbiting an A-type star
($P = 3.3$~d), and Breton et al.\ (2012) discovered a WD 
and F-star binary system, KOI-1224 ($P = 2.7$~d). 
Two more systems orbiting A-type stars were recently discovered by 
Rappaport et al.\ (2015), KIC~9164561 ($P=1.3$~d) and KIC~10727668 ($P=2.3$~d). 
Maxted et al.\ (2011, 2014) reported on the detection of 18 similar 
short-period systems ($P=0.7 - 2.2$~d) in photometric time series from 
the ground-based {\it Wide Angle Search for Planets} survey program. 
Photometric detection of longer period systems is less probable 
because the orbital inclination must be extremely close to $90^\circ$
for transits to occur.  However, longer period systems can be found through 
extensive, time series, radial velocity measurements from spectroscopy. 
For example, Gies et al.\ (2008) used radial velocity measurements 
to show that the nearby star Regulus is a spectroscopic binary ($P=40.1$~d)
consisting of a B-star with a probable WD companion. 
These hot companions of main sequence stars are probably related 
to the subdwarf companions of rapidly rotating Be stars that are 
detected through ultraviolet spectroscopy (Peters et al.\ 2013).  

Close binaries are common among intermediate mass stars, and many of these will 
experience large scale mass transfer (Rappaport et al.\ 2009; van Kerkwijk et al.\ 2010; 
Di Stefano 2011; Clausen et al.\ 2012).  As the initially more massive star 
grows in radius, this donor star will begin Roche lobe overflow (RLOF)
and transfer mass and angular momentum to the companion (the mass gainer).  
The orbit will shrink until the masses are equal, and if contact 
can be avoided, then additional mass transfer will cause an 
expansion of the orbit that ceases once the donor has lost its 
envelope.  The resulting system will consist of a stripped-down and
hot donor star in orbit around a rapidly rotating and more massive
gainer star.  

Detecting binaries in this stage of evolution is difficult because 
the small donor stars are relatively faint and lost in the glare 
of the brighter companions.  Furthermore, the donors are now low mass objects
that create only small reflex orbital motions in the gainer stars, 
so detection through spectroscopy is challenging. 
Thus, the {\it Kepler} discoveries offer us a remarkable opportunity
to investigate the properties of stars that are known examples of
this post-mass transfer state.  The hot companions will contribute 
a larger fraction of the total flux at shorter wavelengths, so a 
direct search for the flux and spectral features associated with 
the hot companion is best done in the ultraviolet.  Indeed, it was 
through UV spectroscopy from the {\it International Ultraviolet 
Explorer} satellite (Thaller et al.\ 1995) and from the {\it Hubble Space 
Telescope} (Gies et al.\ 1998) that the hot subdwarf companion of
the Be star $\phi$~Per was first detected.   

Here we report on an UV spectroscopic investigation of KOI-81
(KIC~8823868; TYC~3556-3094-1; 2MASS~J19350857+4501065)  
made possible with the {\it HST} Cosmic Origins Spectrograph (COS) that 
has revealed the spectral features of the hot companion for the first time. 
We describe the COS observations and supporting ground-based 
spectroscopy in \S2.  We present radial velocity measurements 
and a double-lined orbital solution in \S3 to obtain mass estimates. 
We then use the derived radial velocity curves to perform a 
Doppler tomographic reconstruction of the component spectra, 
and we compare the reconstructed spectra to model spectra to 
derive effective temperatures and projected rotational velocities (\S4). 
The {\it Kepler} light curve outside eclipses shows evidence 
of pulsational and rotational frequency signals that we
discuss in \S5.  The transit light curve is analyzed in \S6 
using a model for the rotationally distorted B-star.  We summarize and 
consider some consequences of the results in \S7. 

 
\setcounter{footnote}{4} 
 
\section{Spectroscopic Observations}                

The Cosmic Origins Spectrograph (COS) is a high dispersion instrument 
designed to record the UV spectra of point sources
(Osterman et al.\ 2011; Green et al.\ 2012). 
The {\it HST}/COS observations of KOI-81 were obtained over
five visits between 2011 June and 2011 October.  These were 
scheduled so that two occurred during each of the quadrature 
phases near the Doppler shift extrema, and the fifth observation
was made during the hot star occultation phase in order to isolate 
the flux of the B-star alone.  The UV spectra were made with 
the G130M grating to record the spectrum over the range from 
1150 to 1450 \AA\ with a spectral resolving power of 
$R=\lambda/\triangle\lambda=18000$.  There are two COS detectors
that are separated by a small gap, so the spectra were made 
at slightly different central wavelengths in order to fill in 
the missing flux: 
1300~\AA\ (four exposures of 447~s), 
1309~\AA\ (three exposures of 399~s), and 
1318~\AA\ (three exposures of 399~s). 
This sequence required an allocation of two orbits for each visit.  
The observations were processed with the standard COS pipeline 
to create wavelength and flux calibrated spectra as {\tt x1d.fits} files 
for each central wavelength arrangement (Massa et al.\ 2013).  
These ten sub-exposures were subsequently merged onto a single 
barycentric wavelength grid using the IDL procedure 
{\tt coadd\_x1d.pro}\footnote{http://casa.colorado.edu/$^\sim$danforth/science/cos/coadd\_x1d.pro}
(Danforth et al.\ 2010).  We created a list of the sharp interstellar
lines in the spectrum, and we cross-correlated each of these spectral 
regions with those in the average spectrum in order to make small 
corrections to the wavelength calibration.  Then the interstellar 
lines were removed in each spectrum by linear interpolation across their 
profiles.  Finally, all five spectra were 
transformed to a uniform grid with a $\log \lambda$ pixel 
spacing equivalent to a Doppler shift step size of 2.60 km~s$^{-1}$
over the range from 1150 to 1440 \AA .  The coadded spectra have
a signal-to-noise ratio of S/N = 110 in the central, best exposed parts. 

We also obtained three sets of complementary ground-based spectra 
of KOI-81.  The first set consists of 19 high dispersion spectra made 
with Tillinghast Reflector Echelle Spectrograph
(TRES\footnote{http://tdc-www.harvard.edu/instruments/tres/})
mounted on the 1.5~m Tillinghast telescope at the Fred Lawrence Whipple Observatory 
at Mount Hopkins, Arizona.  These spectra cover the optical range from 
3850 to 9100 \AA\ with a resolving power of $R=\lambda/\triangle\lambda =48000$.  
The spectra were processed and rectified to intensity 
versus wavelength using standard procedures (Buchhave et al.\ 2010), and 
they are available at the {\it Kepler} Community Follow-up Observing Program 
(CFOP) website\footnote{https://cfop.ipac.caltech.edu/home/}.
A second set of six moderate resolution spectra were made in 2010 and 2012 
with the Kitt Peak National Observatory (KPNO) 4~m Mayall telescope and the
Ritchey-Chretien (RC) Focus Spectrograph.  These were made with the 
BL~380 grating (1200 grooves mm$^{-1}$) to record the spectrum 
between 3950 to 4600 \AA\ with a resolving power of $R=6300$. 
Finally a third set of five, lower resolution, and flux calibrated spectra 
were obtained in 2010 with the Mayall telescope and RC spectrograph using the 
KPC-22B grating (632 grooves mm$^{-1}$) to cover the region from 
3577 to 5058 \AA\ with a  resolving power of $R=2500$.  
This third set is also available at the CFOP website. 

 
\section{Radial Velocities and Orbital Elements}    

Our first goal was to measure the orbital motion of the primary
star using the COS UV spectra.  Direct inspection of the spectra 
showed that the main features were very broadened and blended 
due to large rotational broadening (\S4).   Consequently, 
we decided to measure the individual spectrum velocities by 
cross-correlating them with a spectral template. 
We formed cross-correlation functions (CCFs) using the hot star 
occultation phase spectrum (made on HJD 2455775.9626)
as the CCF template.  The calculation was made using only 
the regions from 1270 to 1300 \AA\ and from 1314 to 1437 \AA\  
(see Fig.~4 below) in order to avoid the low wavelength 
region where the secondary's flux becomes larger and to remove
the \ion{Si}{2} $\lambda\lambda 1260, 1264, 1304, 1309$ features that 
may be affected by incomplete removal of the interstellar components. 
The resulting CCFs are extremely broad, 
and we found the center of each CCF by measuring the wing 
bisector position by convolving the CCF wings with oppositely 
signed Gaussian functions (Shafter et al.\ 1986).  
We similarly measured the center of the CCF of the hot star occultation 
phase spectrum with a model spectrum for the B-star (\S4) from the UVBLUE 
library (Rodr\'{i}guez-Merino et al.\ 2005), and this offset was added 
to the relative velocities to transform them to an absolute scale. 
The resulting radial velocities are collected in Table~1 that
lists a leading P or S for primary B-star or secondary hot star 
(see below), the heliocentric Julian date of mid-observation, 
the orbital phase based upon the {\it Kepler} solution for the 
time of mid-transit (B-star superior conjunction), the radial velocity, 
the difference between the radial velocity and the systemic velocity
for the specific observation set (see below),  
the uncertainty $\sigma$, the observed minus calculated ($O-C$) velocity 
residual from the fit (\S4), and the source (HST/COS in this case). 
Both pairs of quadrature observations were separated in time 
by only a few hours and are expected to have almost the same orbital 
velocity, so we used the absolute differences of the 
measured velocities of these pairs to estimate $\sqrt{2} \sigma$.

\placetable{tab1}      
  
We realized at the outset that finding and measuring the spectral 
lines of the companion would be challenging because of its relative
faintness.  However, detection is favored at the shorter wavelength 
part of the UV spectrum because the hotter companion contributes 
relatively more flux there.  An inspection of the spectra in the 
region with $\lambda < 1200$~\AA\ did indeed
hint at the presence of a velocity variable, narrow-lined component. 
We needed to isolate the flux contribution from the bright B-star 
in this region in order to remove its flux from each spectrum and
reveal the secondary's spectral lines in the flux difference. 
This was accomplished using the hot star occultation phase 
spectrum to represent the flux of the B-star alone. 
This spectrum was smoothed by convolution with a Gaussian function of 
FHWM = 133 km~s$^{-1}$ in order to increase the S/N ratio
without unduly altering the spectral shape.  Then the smoothed 
version was shifted in velocity according to the predicted orbital motion of
the primary star, and a difference spectrum was formed for each of the four 
quadrature phase observations.  Renormalized and averaged versions of 
these quadrature phase spectra are illustrated in Figure~1, and these 
show a host of weak and narrow lines that are Doppler shifted 
as expected for the orbital motion of the hot secondary.  
The flux scaling for the hot companion is not accurate because of 
the rudimentary means of the B-star's flux removal, but it is 
sufficient to reveal the lines of the hot component so that 
its radial velocity can be measured.  We did so by calculating
the CCF of each quadrature spectrum using a hot model spectrum (\S4) 
for the template, and the center of the narrow signal in the resulting 
CCF was estimated by fitting a parabola to its peak (with uncertainties
determined using the method of Zucker 2003).  The derived radial
velocities of the secondary are listed at bottom of Table~1 in 
rows with a leading column marked S.  No measurement is given 
for the occultation phase spectrum because the flux of the hot 
component is totally blocked then (Rowe et al.\ 2010).  

\begin{figure}
\begin{center} 
{\includegraphics[angle=90,height=12cm]{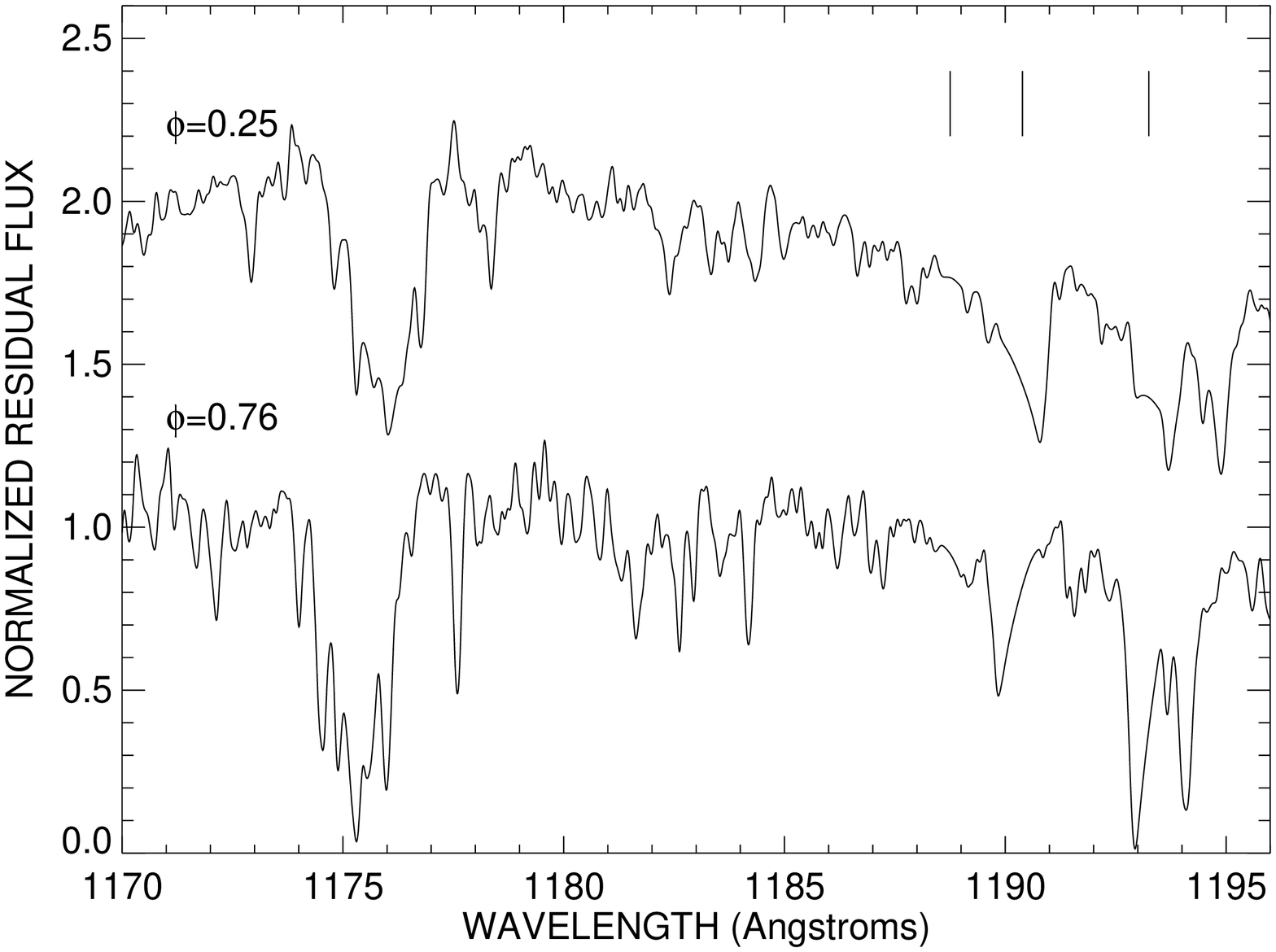}}
\end{center} 
\caption{Difference spectra of KOI-81 formed by subtracting 
the occultation phase spectrum of the primary alone. 
The sharp absorption lines from the subdwarf star appear near maximum 
redshift (blueshift) at orbital phase $\phi=0.25$ ($\phi=0.76$).
The vertical line segments indicate the positions where 
interstellar lines were removed from the spectrum.
\label{fig1}} 
\end{figure}

\placefigure{fig1}     

We also measured radial velocities for the primary B-star using
the higher resolution, ground-based spectra.  The 19 TRES spectra 
have a S/N that is too low for measurement of the weak and 
broadened He and metallic lines, but they are sufficient 
to measure the Doppler shifts in the strong and broad, 
H Balmer lines.  We formed CCFs of each of 
H$\alpha$, H$\beta$, H$\gamma$, H$\delta$, and H$\epsilon$ 
with model template spectra from the 
BLUERED\footnote{http://www.inaoep.mx/$^\sim$modelos/bluered/go.html} 
grid (Bertone et al.\ 2008).  The radial velocity was estimated
by fitting a parabola to the peak of the CCF for each line. 
The quality of the measurement varies depending on the position of
the feature relative to the echelle blaze function (best for 
H$\alpha$ and H$\gamma$ which fall close to the blaze maximum), 
so we formed weights for each of the Balmer lines proportional 
to the inverse square of the standard deviation of the velocity 
measurements.   We then determined the weighted mean and standard
deviation of the mean from the set of five line measurements for 
each observation, and these are listed in Table~1 
(noted by TRES in the final column). 

We followed a similar procedure to measure radial velocities 
for the six KPNO moderate resolution spectra.  
We calculated CCFs using a model B-star template (from the UVBLUE grid)
and including all the lines in the region from 4050 to 4520 \AA\ 
(dominated by H$\gamma$ and H$\delta$), and we fit a parabola
to the CCF peak to estimate radial velocities.  These are 
identified in Table~1 by KPNO in the final column.   

We fit orbital elements for each of the primary and secondary velocity 
sets using the program described by Morbey \& Brosterhus (1974). 
We fixed the orbital period and epoch of mid-transit to those 
derived by the {\it Kepler} project (dated 2014 November 24) 
that are posted at the CFOP website.  Because the transit and
occultation light curves have the same duration and are separated 
by half of the orbital period, we assumed that the orbit is circular. 
We set the fitting weights to the inverse square of the uncertainties. 
Because the COS, TRES, and KPNO measurements are derived from differing 
spectral ranges and instruments, we might expect that there will 
be systematic differences in the velocities for each set. The upper panel of 
Figure~2 shows the measured radial velocities from Table~1, using different
symbols for each set of spectra, as well as preliminary fits of each set,
and there do indeed appear to be constant offsets between the sets. 
We found independent estimates for the systemic velocity $\gamma_i$ of each set by 
iterating between global (all data) and individual sets of velocities 
for the orbital fits.  This was done by fixing the semiamplitude $K_1$
from the global fit to find estimates of $\gamma_i$ for solutions to each set.
Then a new global fit was made using velocity differences (subtracting the
$\gamma_i$ value associated with each set) to find a revised $K_1$. 
This procedure converged after a few iterations to the final adopted
fit presented in Table~2.  The $\gamma_i$-corrected velocities 
(listed under column $V_r - \gamma_i$ in Table~1) and $2\sigma$ 
range of the fitted velocity of the primary are shown in the lower panel
of Figure~2, and the combined velocity curves of the primary and secondary 
are illustrated in Figure~3.   We also made a fit of the $\gamma_i$-corrected 
velocities of the primary using a Markov Chain Monte Carlo algorithm that led 
to almost the same result, $K_1 = 6.75^{+0.43}_{-0.44}$ km~s$^{-1}$.  
However, we conservatively adopt the somewhat larger uncertainty estimates 
from the non-linear least squares program in what follows.
Van Kerkwijk et al.\ (2010) estimated 
$K_1 \approx 7$ km~s$^{-1}$ from the Doppler boosting in the light 
curve of KOI-81, and their result is verified through our direct 
Doppler shift measurement of $K_1 = 6.7 \pm 0.7$ km~s$^{-1}$.

\placetable{tab2}      
  
\begin{figure} 
\begin{center} 
 {\includegraphics[angle=0,height=12cm]{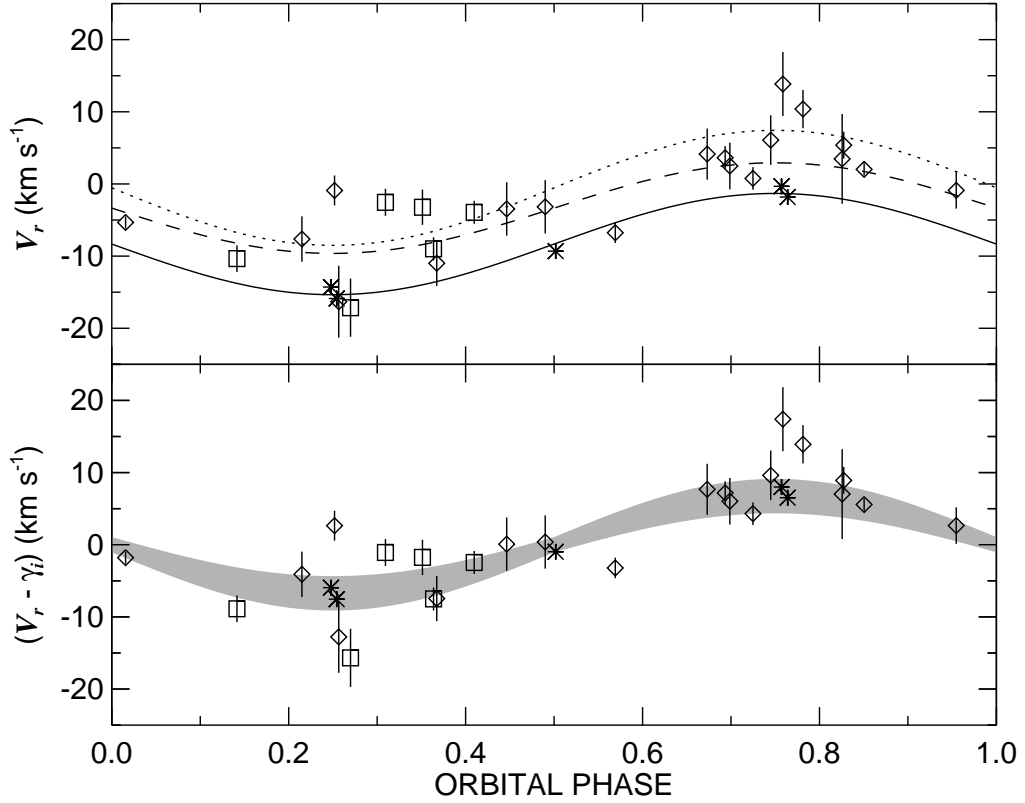}} 
\end{center} 
\caption{Top panel: The radial velocity data from Table~1 for the primary star
plotted as a function of orbital phase.  Phase 0.0 corresponds to the time of 
central transit of the hot subdwarf across the face of the brighter B-type star.   
The asterisk, diamond, and square symbols represent measurements from 
COS, TRES, and KPNO, respectively, and the solid, dashed, and dotted lines
show preliminary circular fits for the same three sets 
(with $K_1 = 7.0$, 6.3, and 8.0 km~s$^{-1}$
and $\gamma_i = -8.3$, $-3.4$, and $-0.5$ km~s$^{-1}$, respectively).  
Lower panel: The radial velocity differences 
(formed by subtracting the systemic velocities for each set reported 
in Table 2) as a function of phase.  The shaded region indicates the $\pm 2 \sigma$
range in the velocity curve for the adopted fit of the combined measurements. 
\label{fig2}} 
\end{figure}

\placefigure{fig2}     

\begin{figure} 
\begin{center} 
{\includegraphics[angle=90,height=12cm]{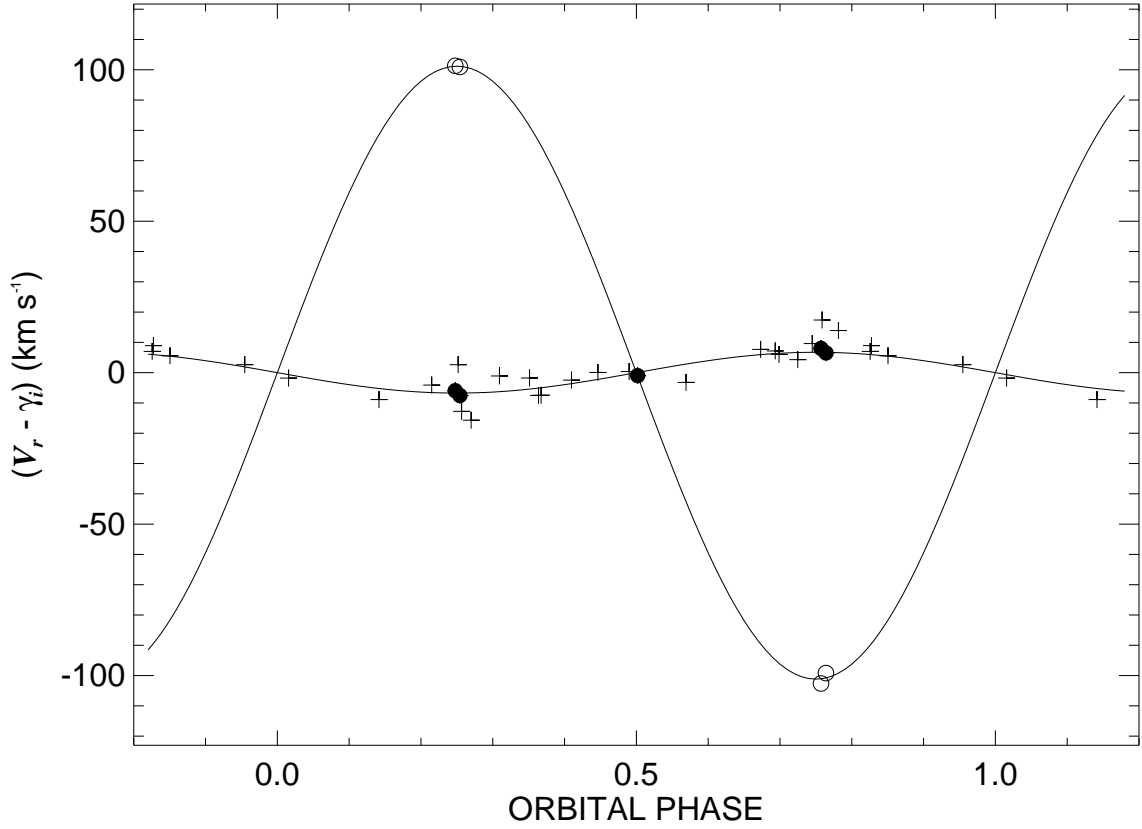}} 
\end{center} 
\caption{Radial velocity curves for KOI-81 and its companion. 
The solid circles and open circles represent the radial velocities 
derived from COS spectra for the B-star and hot subdwarf, respectively.
Plus signs represent the measurements of the B-star velocity from 
ground-based spectroscopy (see Fig.~2).  
\label{fig3}} 
\end{figure}

\placefigure{fig3}     

The orbital inclination is very close to $i=90^\circ$.  We derive a 
value from the transit light curve (\S6) of $i=88\fdg97 \pm 0\fdg04$, 
which is intermediate between the estimates from 
Rowe et al.\ (2010) of $i=88\fdg2 \pm 0\fdg3$ and 
from the {\it Kepler} project posted at CFOP of $i=89\fdg95$.  
Thus, we can use our estimate to derive the physical masses
from the $M_1 \sin^3 i$, $M_2 \sin^3 i$ products, and these 
are given in Table~3.  Furthermore, the average density $\rho$ 
of the B-star can be directly estimated from the transit light curve
(provided the star is spherical; see \S6), and the {\it Kepler} 
project finds $\rho=0.280 \pm 0.005$ g~cm$^{-3}$ for KOI-81
(reported at the CFOP website).  We used this value together with 
the mass estimate $M_1$ to arrive at the radius $R_1$ reported in 
Table~3.  Finally, the CFOP website gives the ratio of the radii derived 
from the transit light curve, $R_2/R_1 = 0.03725 \pm 0.00026$,  
and we used this ratio to find $R_2$ (Table~3).  Table~3 also 
presents the gravitational acceleration $\log g$ derived from the 
mass and radius information.  We must examine the spectrum 
of the system to derive the remaining stellar parameters. 

\placetable{tab3}      
 
 
\section{Tomographic Reconstruction of the UV Spectra}  
 
We used a Doppler tomography algorithm (Bagnuolo et al.\ 1994)
to extract the individual UV spectra of the primary and secondary stars. 
This is an iterative scheme that uses estimates of the orbital Doppler
shifts of each component and their flux ratio to derive 
reconstructed spectra for both stars.  We initially assumed 
featureless continua as the starting approximation for the 
spectra of both stars, but after comparison with model spectra, 
we ran the algorithm again using the models as starting values, 
and this choice helped to limit the continuum wander in the 
resulting spectral reconstructions.  We present in Figures 4 and 5 
the reconstructed UV spectra derived from the four COS spectra 
obtained at the quadrature phases.  Figure~4 also shows the 
excellent agreement between the reconstructed spectrum of 
the primary and the hot star occultation phase spectrum 
that represents the flux of the primary alone.  Figures 4 and 5
indicate some of the principal lines as well as the locations 
of the interstellar lines, where their removal by interpolation 
may have interfered with the accurate reconstruction of the spectra. 
All the spectra in Figures 4 and 5 are smoothed by convolution with a 
boxcar function of width 133 km~s$^{-1}$ in order to reduce the noise
and facilitate intercomparison of the line features. 

\begin{figure} 
\begin{center} 
{\includegraphics[angle=90,height=12cm]{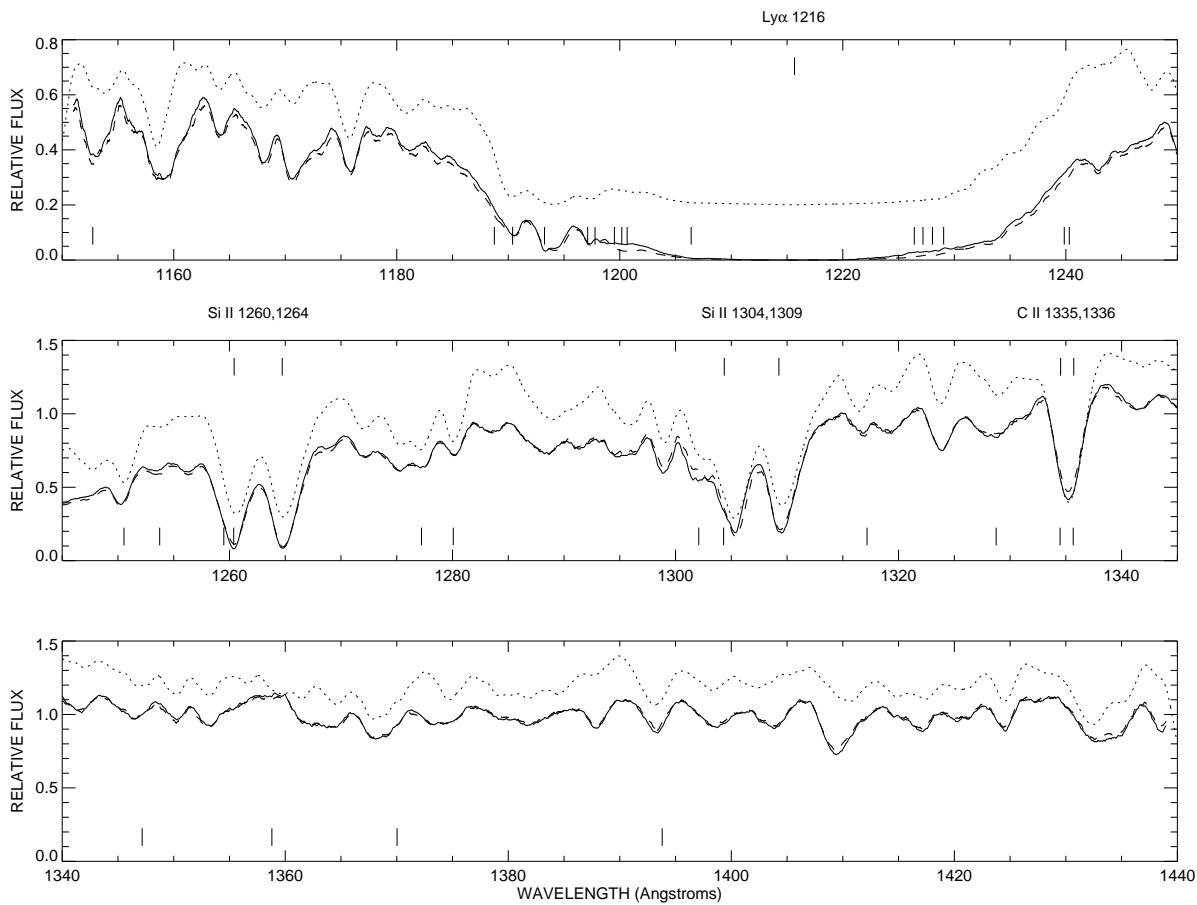}} 
\end{center} 
\caption{The reconstructed UV spectrum of the B-star (solid line) 
depicted in three successive wavelength panels.  The dashed line
shows the COS spectrum obtained when the subdwarf was occulted, 
and it is identical within uncertainties with the reconstructed 
spectrum.  A model spectrum constructed from the UVBLUE grid 
appears as a dotted line above the reconstructed spectrum 
(offset by $+0.2$ in units of normalized flux).  Several strong 
absorption lines are identified and marked by the upper vertical 
line segments, while the lower line segments indicate the locations 
where interstellar lines were removed from the spectrum. 
\label{fig4}} 
\end{figure}

\placefigure{fig4}     
 
\begin{figure} 
\begin{center} 
{\includegraphics[angle=90,height=12cm]{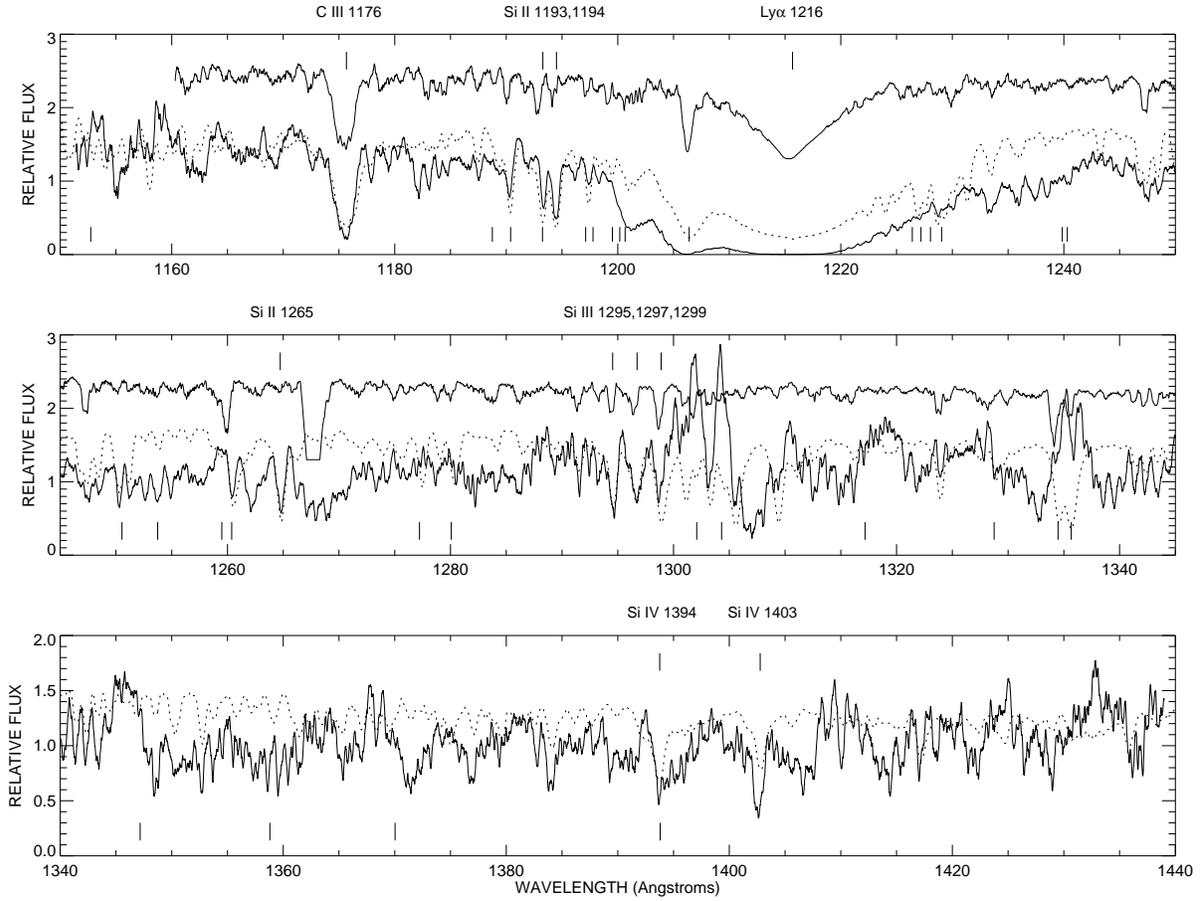}} 
\end{center} 
\caption{The reconstructed UV spectrum of the subdwarf star depicted 
in three successive wavelength panels (in the same format as Fig.~4). 
The top two panels also show for comparison the spectrum of the 
hot subdwarf CPD$-64^\circ 481$ (solid line offset by $+1.3$ in 
normalized flux). 
\label{fig5}} 
\end{figure}

\placefigure{fig5}     

We compared the reconstructed spectra with model spectra 
from the UVBLUE grid of high resolution 
spectra\footnote{http://www.inaoep.mx/$^\sim$modelos/uvblue/go.html}
calculated by Rodr\'{i}guez-Merino et al.\ (2005) that are 
based upon the ATLAS9 model atmosphere code and SYNTHE 
radiative transfer code developed by R.\ L.\ Kurucz. 
We used their solar metallicity models that incorporate a 
microturbulent velocity of 2 km~s$^{-1}$. 
We made simple bilinear interpolations of flux in the 
$(T_{\rm eff}, \log g)$ plane to derive the model spectra. 
We adopted $\log g_1 = 4.13$ for the primary (Table~3),
but set $\log g_2 = 5.0$ for the secondary because this 
value is the largest available in the UVBLUE grid.  
The model spectra were rebinned onto the observed wavelength grid
and then convolved with the instrumental broadening function 
(from the COS line spread 
function\footnote{http://www.stsci.edu/hst/cos/performance/spectral\_resolution/fuv\_130M\_lsf\_empir.html}
for a central wavelength of 1300 \AA ) and with a 
rotational broadening function (using linear limb darkening 
coefficients from Wade \& Rucinski 1985). 

We first compared the reconstructed and model spectra to estimate 
the projected rotational velocity $V \sin i$ of each star.
This was done by forming a $\chi^2$ goodness-of-fit statistic 
between the observed and model spectra over a test grid of 
$V \sin i$ values, and then finding the $V \sin i$ corresponding 
to the minimum $\chi^2$.  This was repeated for a series of 
wavelength regions that contained well defined absorption lines 
or line blends, and the mean and standard deviation of the 
derived $V \sin i$ for the primary is presented in Table~3. 
The primary B-star is indeed a very rapidly rotating star, 
a property noted first by van Kerkwijk et al.\ (2010).  
The lines of the hot secondary companion, on the other hand, 
appear very sharp and any rotational broadening is unresolved 
in the COS spectra, so we present only an upper limit in Table~3. 

Next we needed to adjust the model spectra for the variable 
interstellar extinction across the COS wavelength band. 
We derived the interstellar reddening by comparing the 
available observed fluxes with a model flux distribution 
for the binary transformed using the extinction law presented 
by Fitzpatrick (1999).  The observed fluxes are shown in 
a spectral energy distribution plot in Figure~6.  These include
rebinned COS measurements, a GALEX\footnote{http://galex.stsci.edu/GR6/} 
NUV measurement\footnote{KOI-81 appears as two sources in GALEX, but we simply summed
the fluxes of what appears to be the trailed image of a single object.}, 
rebinned KPNO low dispersion spectrophotometry, and 2MASS 
and WISE photometric fluxes.  The model fluxes were a sum of 
Kurucz models for the primary and secondary that were attenuated
according to the extinction law for a ratio of total-to-selective
extinction of $R=3.1$.  The fit shown in Figure~6 was made with 
a reddening of $E(B-V) = 0.169 \pm 0.008$ mag and a limb darkened
angular diameter of the primary of $\theta = 0.0182 \pm 0.0005$ milliarcsec.
The former agrees with prior estimates ($E(B-V)= 0.175$ and 0.193 mag 
in the {\it Kepler} Input Catalog and the GALEX Catalog, respectively), 
and the latter indicates a distance of 1.25 kpc (using $R_1/R_\odot$ from Table~3). 
The extinction in the COS band was calculated for the derived value
of $E(B-V)$, and each model was multiplied by this extinction curve
to account for the greater interstellar attenuation at lower wavelength. 

\begin{figure} 
\begin{center} 
{\includegraphics[angle=90,height=12cm]{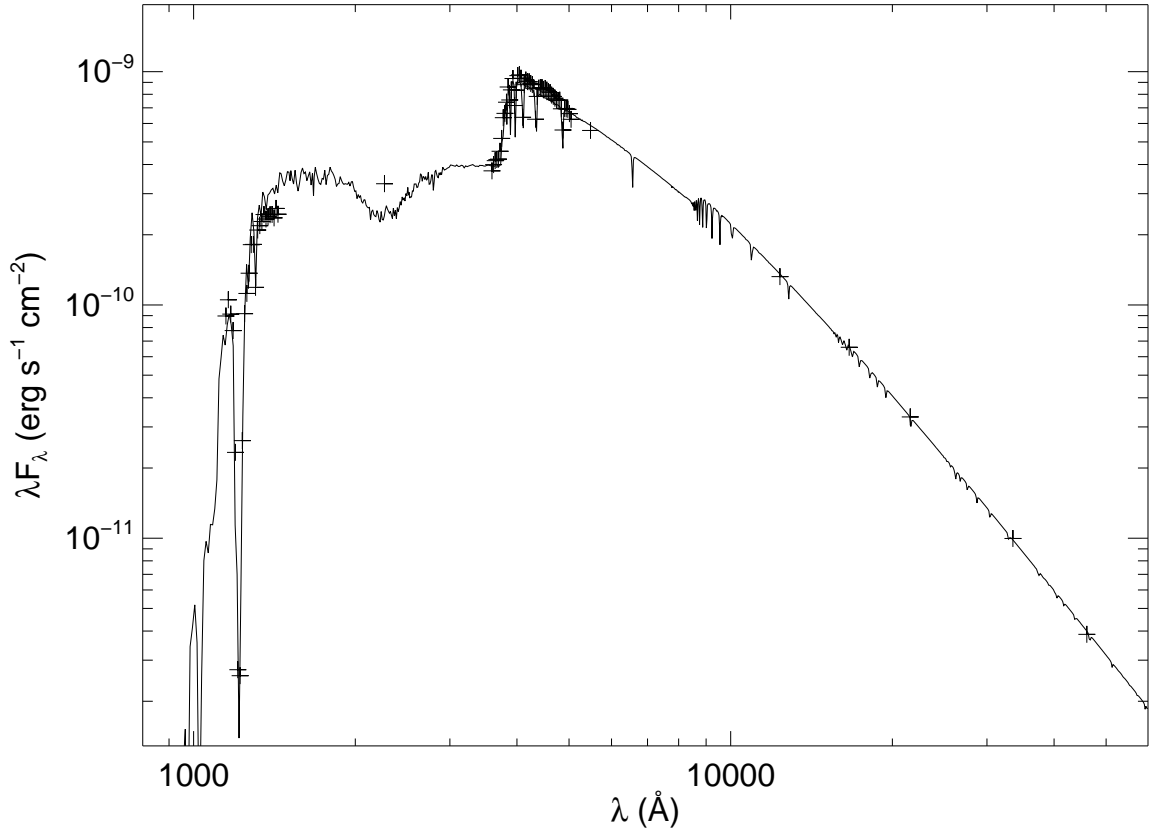}} 
\end{center} 
\caption{The spectral energy distribution of the KOI-81 binary 
including (from short to long wavelength) fluxes from COS (UV), GALEX (NUV), 
the KPNO 4~m (optical, low resolution spectroscopy), 2MASS (near-IR), 
and WISE (mid-IR).  The solid line depicts a Kurucz atmosphere 
flux model for the combined system that is based upon a fit of 
the reddening and the B-star angular diameter.  
\label{fig6}} 
\end{figure}

\placefigure{fig6}     

With the gravity, projected rotational velocity, and interstellar extinction set, 
we then calculated model spectra over a grid of effective temperature 
and determined a $\chi^2$ goodness-of-fit as a function of $T_{\rm eff}$.
The best fit temperatures and their estimated uncertainties from this 
comparison of the reconstructed and model spectra are listed in Table~3. 
These uncertainties do not account for possible differences from the  
assumed microturbulent velocity and abundance.  Nevertheless, these models 
offer a framework to interpret the appearance of the UV spectra.   

Figure~4 shows the derived model spectrum offset above the 
reconstructed spectrum of the primary star.  The overall agreement
is very good in both the appearance of the lines and line blends. 
The \ion{C}{2} $\lambda\lambda 1335, 1336$ feature is weaker than 
predicted, but this may result from our means of removal of interstellar
line components in this feature.  
The flux in the optical wavelength range is totally dominated by 
the light from the B-star primary (\S6), so we did not attempt
Doppler tomography reconstructions of the optical spectrum. 
However, we show in Figure~7 two optical lines of special interest. 
The most temperature sensitive feature in 
the optical spectrum at $T_{\rm eff}\approx 10$~kK is the \ion{Ca}{2}
$\lambda 3933$ K line that grows rapidly in strength with decreasing 
temperature (Gray \& Corbally 2009).  The average TRES spectrum of this feature is 
shown in Figure~7 ({\it left}) together with model spectra for $T_{\rm eff}=11.7$~kK
and 10.6~kK, and a comparison with the observed K-line suggests that 
the effective temperature may be somewhat lower than that derived from the 
ultraviolet spectrum (but still within the uncertainties). 
On the other hand, the difference may result from gravity darkening in 
the rapidly rotating B-star that creates a cooler equatorial region (\S6) 
that would promote the appearance of lines favored in cooler atmospheres.
Figure 7 also illustrates the average observed and model profiles of 
the gravity sensitive H$\gamma$ line ({\it right}), and the observed profile
has a shape that is approximately consistent with predictions for 
the adopted gravity of the primary. 
 
\begin{figure} 
\begin{center} 
{\includegraphics[angle=90,height=6cm]{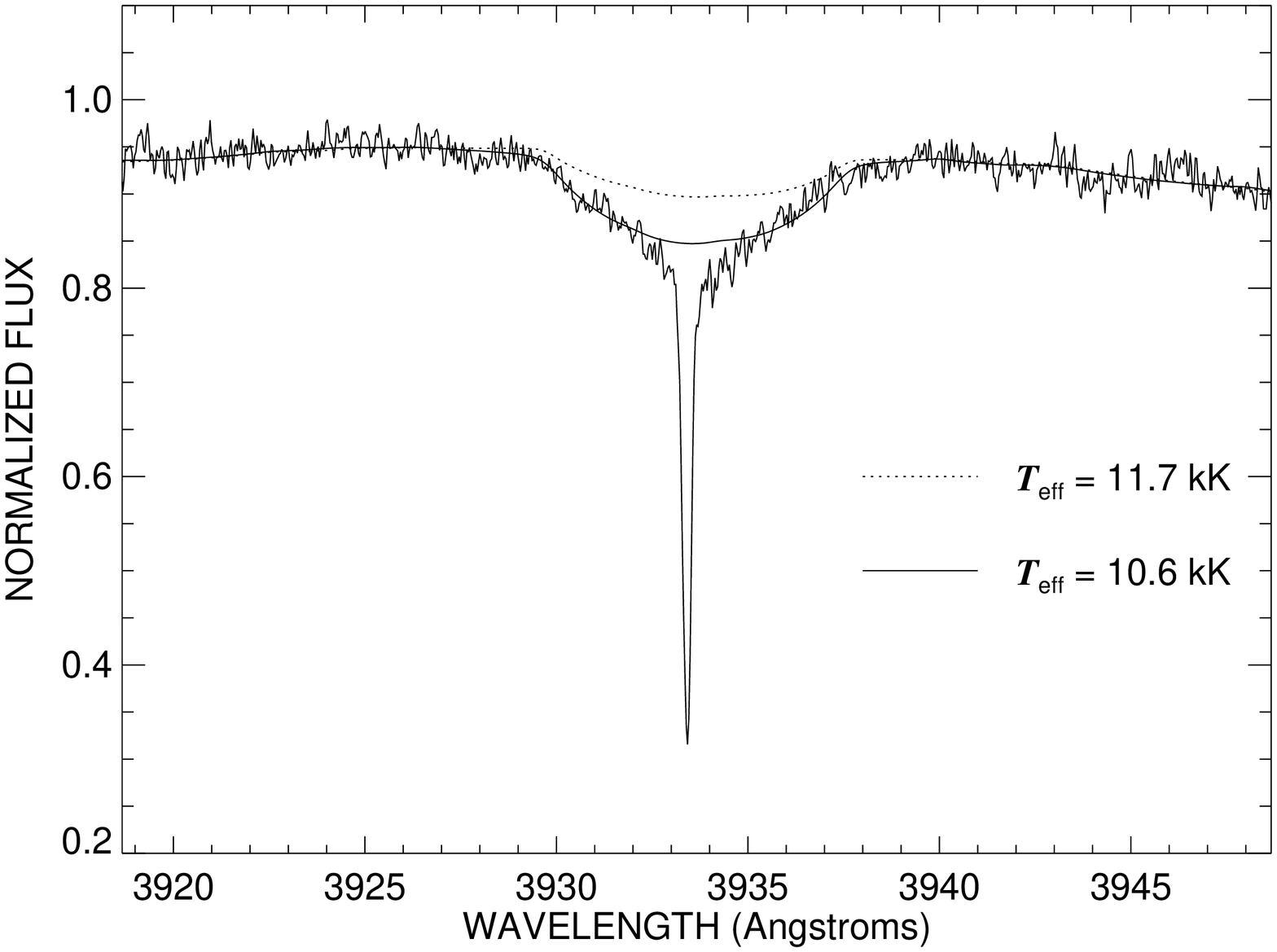}} 
{\includegraphics[angle=90,height=6cm]{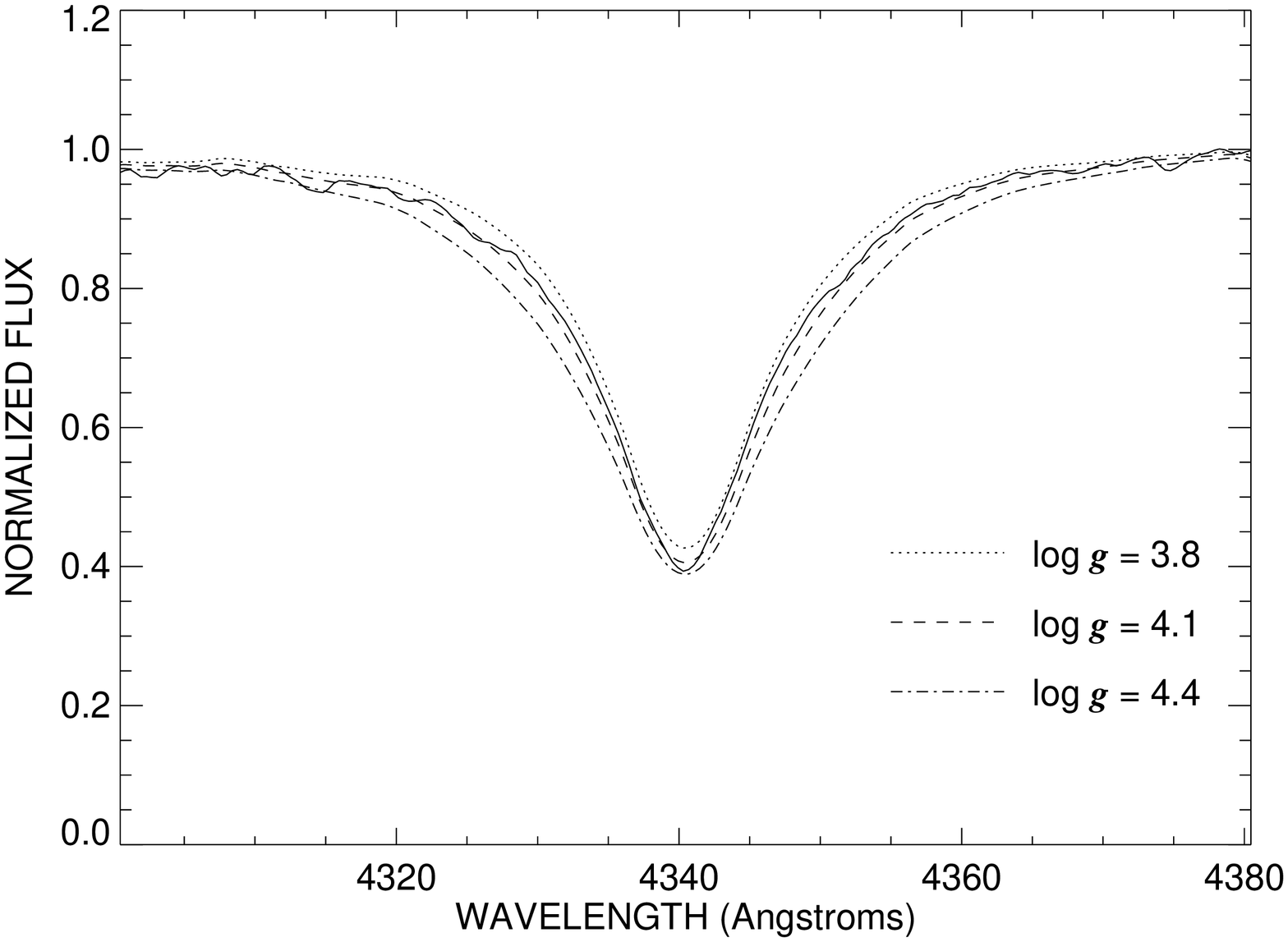}} 
\end{center} 
\caption{
{\it Left:} 
The \ion{Ca}{2} $\lambda 3933$ feature (structured solid line) from the average
of the TRES high resolution spectra.  A sharp, interstellar component
(with a radial velocity of $V_r = -19.2 \pm 0.9$ km~s$^{-1}$) is situated
near the center of the broad photospheric component.  The solid and dotted 
lines show UVBLUE model profiles for $T_{\rm eff}= 10.6$ and 11.7 kK, respectively.
{\it Right:} 
The H$\gamma$ line profile (solid line) from an average 
of the KPNO 4~m moderate resolution spectra.  The other lines
represent UVBLUE model spectra for three choices of surface gravity.  
\label{fig7}} 
\end{figure}

\placefigure{fig7}     

The reconstructed secondary spectrum shown in Figure~5 is 
quite noisy because this component is relatively faint (see below), 
but there is satisfactory agreement between 
the reconstructed and model spectra in the vicinity of some features. 
For example, the reconstructed and model spectra of the secondary appear similar 
in the lines of \ion{C}{3} $\lambda 1176$, \ion{Si}{2} $\lambda\lambda 1193, 1194$, 
\ion{Si}{3} $\lambda\lambda 1295, 1297, 1299$, and \ion{Si}{4} $\lambda\lambda 1394, 1403$.
The latter are only seen in the spectrum of the hotter secondary. 
However, the agreement is less satisfactory in the regions where 
interstellar lines were removed (near 1229, 1251, 1303, 1335, and 1347 \AA ).
The model spectra were formed using UVBLUE models with $\log g = 5.0$, 
the largest value in the grid, which is smaller than the estimated gravity, 
$\log g = 5.8$.  This has two important consequences.  First, 
the model Ly$\alpha$ line is narrower than observed because the linear Stark 
broadening associated with the lower gravity model is insufficient to match 
the observed broadening.  Second, the effective temperature estimation 
is based upon the relative strengths of transitions corresponding to different
ionization states, and according to the Saha equation, line formation in a 
denser medium (at higher gravity) will shift the ionization balance to less
ionized states.  Consequently, a good fit of the spectrum would also be possible 
with a higher gravity and higher temperature model, so our derived effective
temperature should be regarded as a lower limit.  We also show in Figure~5 an 
example spectrum of the B-subdwarf CPD$-64^\circ 481$ that was made with the 
Space Telescope Imaging Spectrograph and E140H grating.  O'Toole \& Heber (2006)
used this and other spectra to estimate $T_{\rm eff}=27.5$~kK and $\log g =5.6$, 
a value of gravity that is closer to that of the secondary in KOI-81. 
Comparing these similar gravity stellar spectra, we see that transitions 
of \ion{Si}{2} are almost absent in the spectrum of CPD$-64^\circ 481$, 
whereas they are quite strong in the spectrum of the KOI-81 secondary,
so the KOI-81 secondary must be cooler than CPD$-64^\circ 481$. 
Thus, we can place lower and upper limits on the effective temperature 
of the secondary of $19 < T_{\rm eff} < 27$~kK.
 
All the spectra in Figures 4 and 5 were normalized in flux by 
dividing by the mean flux over the range between 1340 and 1400 \AA .
We set the monochromatic flux ratio from this normalization to
$F_2(1370)/F_1(1370)=0.05 \pm 0.02$ based upon a comparison of the
line depths in the reconstructed and model spectra of the secondary. 
This comparison was made using relative line depths in the 
vicinity of the lines that matched reasonably well, because 
there is too much wander in the continuum of the reconstructed 
spectrum of the secondary for a reliable global fit. 
We caution that the model line depths are sensitive to the 
adopted values of temperature, gravity, and especially microturbulence, 
so the flux ratio may need revision if different assumptions are made.

 
\section{Non-orbital Frequencies in the Light Curve} 

Both Rowe et al.\ (2010) and van Kerkwijk et al.\ (2010) noted that 
the light curve of KOI-81 shows evidence of pulsations, and 
it is interesting to consider the pulsational frequencies and 
their relation to the rotational frequency of the B-star primary. 
We calculated the Fourier transform amplitude of the 
detrended, long cadence {\it Kepler} observations from quarters 
0 to 17 using the package Period$04$ (Lenz \& Breger 2005).  
The portions of the light curve covering the transit
and occultation of the secondary were removed from the time series
in order to focus on non-orbital timescales of variation. 
We adopted an empirical noise level by smoothing the envelope 
of the residual spectrum after prewhitening all significant frequencies 
with $S/N >3$ (except for the broad rotational feature; see below).
The uncertainties were calculated following Kallinger et al.\ (2008).
The Fourier spectrum is dominated by a very strong signal with 
a frequency of $f_1=0.722974$ cycles~d$^{-1}$ (period of 1.38318~d),
so we removed this signal by pre-whitening to uncover 
the remaining periodic signals that are plotted in Figure~8.
There are a number of strong signals that appear well above 
the noise level, and the frequencies, amplitudes, sinusoidal phases 
(relative to the epoch of central transit), and peak signal-to-noise
ratio are listed in Table~4.  The final column of Table~4 lists 
several numerical relations among these frequencies, and identifies 
one frequency $f_{12}$ that corresponds to the ellipsoidal (tidal)
variation with half the orbital period (discussed by van Kerkwijk et al.\ 2010). 
 
\begin{figure} 
\begin{center} 
 {\includegraphics[angle=0,height=12cm]{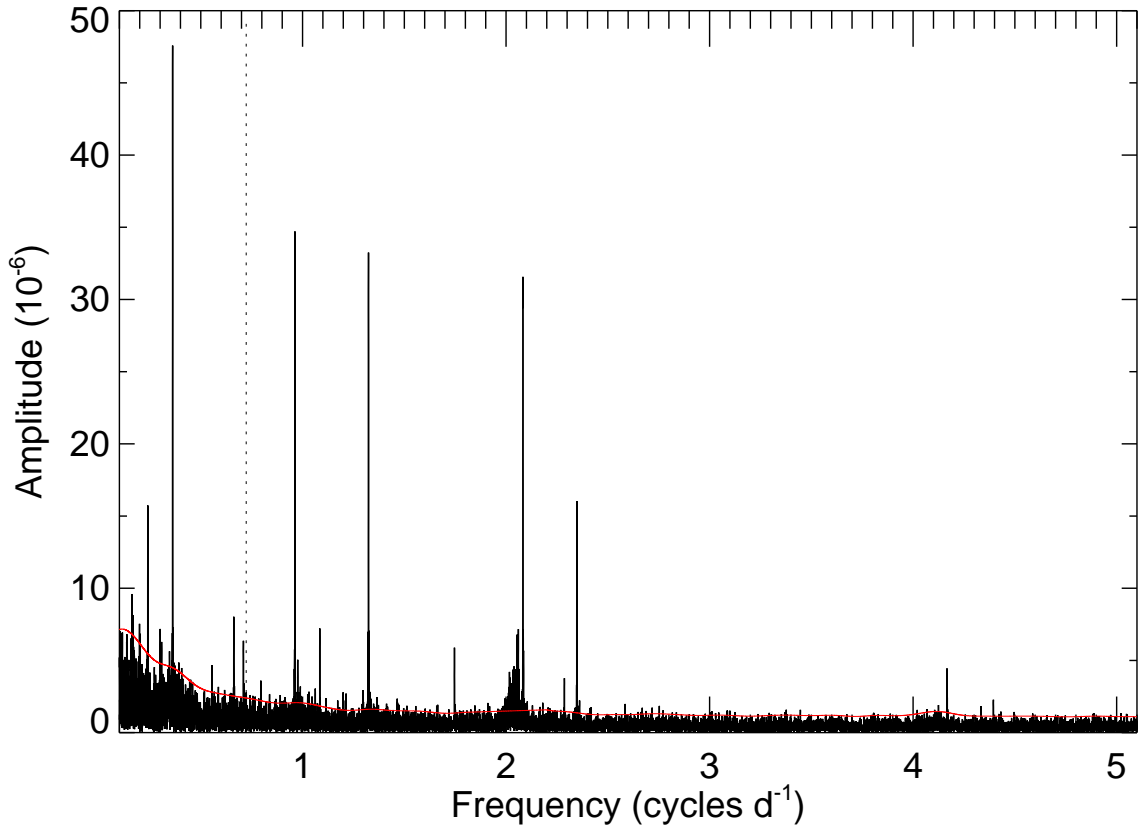}} 
\end{center} 
\caption{The Fourier amplitude spectrum of {\it Kepler} 
short cadence photometry after prewhitening the dominant peak 
at $0.72297$ d$^{-1}$ indicated by the dotted line.  The empirical 
noise level is indicated by the red solid line. 
\label{fig8}}
\end{figure}

\placefigure{fig8}     

\placetable{tab4}      
 
The dominant periodicity (1.38~d) is unrelated to the orbital period,
and it probably represents a strong and long-lived pulsational mode.
This and the other low frequency pulsation signals may be of two possible kinds.  
First, the B-star primary of KOI-81 has a temperature and radius that are similar 
to those of the slowly pulsating B-type (SPB) stars (Pamyatnykh 1999), 
and these stars display relatively long period $g$-mode oscillations.
Second, the B-star is a rapid rotator, and Townsend (2005) and Savonije (2013)
argue that rapidly rotating, late-type B-stars can experience 
retrograde mixed modes of low azimuthal order $m$.  The rotational 
frequency of the B-star is probably $\approx 2$ cycles~d$^{-1}$
(see below), so the smaller frequency of the dominant signal is 
consistent with retrograde nonradial pulsation.  

There is a broad distribution of peaks just above $2$ cycles~d$^{-1}$, 
and we show an enlarged version of the Fourier amplitude in this 
vicinity in the top panel of Figure~9.  A wide distribution appears 
around $2.04$ cycles~d$^{-1}$ that is accompanied by a strong peak 
at $f_5 = 2.08287$ cycles~d$^{-1}$.  The lower panel shows the 
residual peaks after pre-whitening and removal of the strong $f_5$
signal, and this reveals the presence of several other significant 
peaks near $f_5$.  This kind of broad feature with a stronger single 
peak at slightly higher frequency has been detected by Balona (2013, 2014)
in the {\it Kepler} light curves of some $19\%$ of A-type stars.  
Balona argues persuasively that this feature probably corresponds to the stellar
rotational frequency.  In his interpretation, the single strong peak 
corresponds to the equatorial rotational frequency and the wider peak
samples the rotational frequencies at different latitudes in stars 
with differential rotation.  Thus, following this line of argument,
we may tentatively identify $f_5$ as the equatorial rotational frequency
of the B-star in KOI-81, and thus, the rotational period is 0.48~d at 
the equator.  Balona (2014) considers several explanations
for the origin of the photometric variation including pulsation, 
rotational modulation by starspots, and tidal variations induced 
by a co-orbiting exoplanet.   We suspect that in the case of 
KOI-81, any co-orbiting planet would have a short period and 
an orbital plane similar to that of the stellar companion, so that 
we might expect to observe a transit signal in the $f_5$ folded 
light curve, but instead the folded light curve is approximately 
sinusoidal in shape.  We speculate that the rotational signals 
in the light curve of KOI-81 and similar stars may result from 
long-lived vortices (Kitchatinov \& R\"{u}diger 2009) that develop 
in the outer atmospheres of rotating stars due to differential rotation 
(similarly to the spots in the atmosphere of Jupiter). 

\begin{figure} 
\begin{center} 
 {\includegraphics[angle=0,height=12cm]{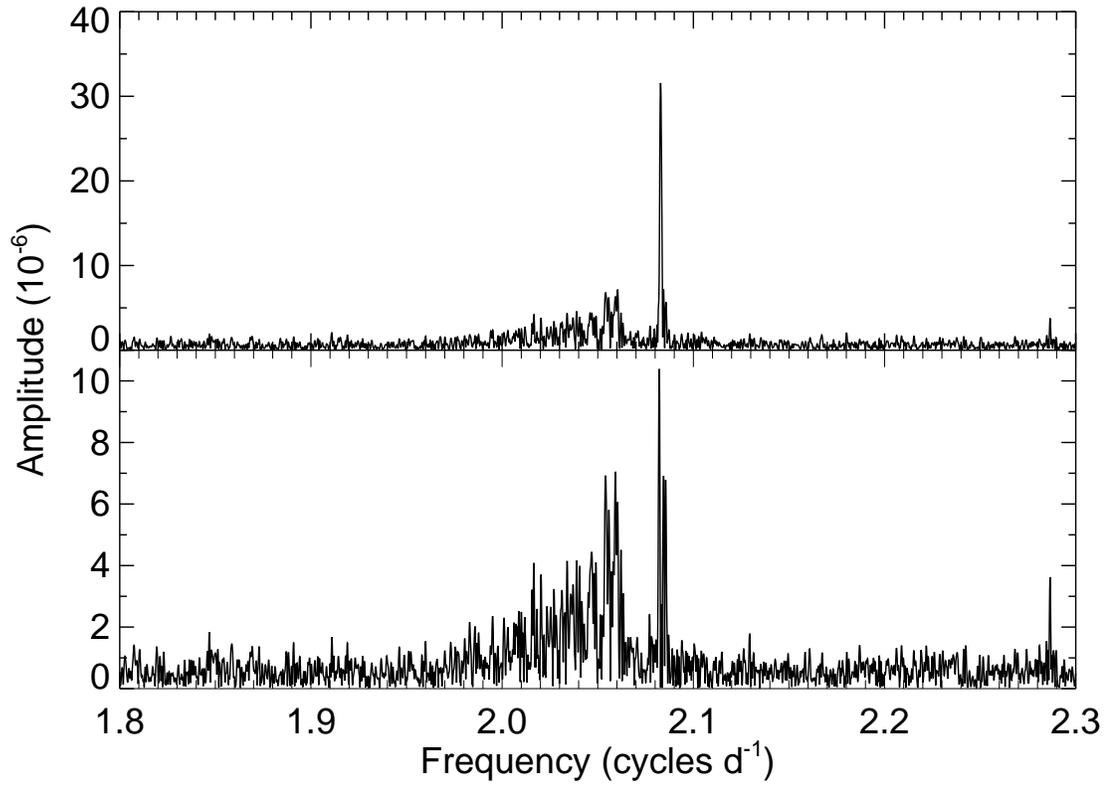}} 
\end{center} 
\caption{Top: The broad differential rotation feature at a frequency 
of about $1.96\sim2.06$ d$^{-1}$ and the adjacent sharp peak 
$f_{5} = 2.08287$ d$^{-1}$. Bottom: the same diagram but with 
the sharp $f_{5}$ peak prewhitened.
\label{fig9}}
\end{figure}

\placefigure{fig9}     

 
\section{Transit Light Curve}                       

The B-star we observe was spun up during the mass transfer process 
to produce a very rapidly rotating star with a rotationally 
broadened spectrum.  It is important to consider how this 
rapid rotation influences our interpretation of the light curve. 
We expect that the spun up star will have a rotational axis 
parallel to the orbital angular momentum vector, so that the 
star's spin axis also has an inclination of $i\approx 90^\circ$. 
We argued in \S5 that the photometric signal $f_5$ is the 
rotational frequency of the B-star.  Then we may estimate the 
star's equatorial radius from 
$R_{1~{\rm equator}} = (V \sin i)/(2 \pi f_5 \sin i) = (2.81 \pm 0.05) R_\odot$. 
This is somewhat larger than what we derived from the mass and 
mean stellar density (Table~3), but is not unexpected for a 
rotationally distorted star in which the equatorial radius 
will be larger (and the polar radius smaller) than the mean radius.

We can model the predicted appearance of the B-star based upon
our estimates of stellar mass, equatorial radius, rotational 
period, and average effective temperature.  We created 
a model image of the specific intensity at a wavelength 
of 6430 \AA , which is the centroid of the {\it Kepler}
instrument response function, using the same methods 
applied in a study of the rapidly rotating B-star Regulus
(McAlister et al.\ 2005).  The star is assumed to have a shape 
that follows the Roche potential for rotation about a point mass, 
and each surface element has a specific intensity defined by 
limb and gravity darkening.  
The specific intensity of a surface element is 
set by interpolation in a set of Kurucz model values 
defined by the local temperature, gravity, and orientation 
of the surface normal to the line of sight.  We found values
of the polar radius and temperature that led to our derived
equatorial radius and surface average temperature.  
An image of the rotationally distorted star appears in Figure~10.
The resulting model has a ratio of 
equatorial to critical velocity (where the gravitational 
and centrifugal accelerations are equal) of 0.74, 
and the radius varies from $2.30 R_\odot$ at the poles 
to $2.81 R_\odot$ at the equator.  The temperature in this 
model varies between 13.8~kK and 10.8~kK from pole to equator, 
and gravity likewise varies from $\log g = 4.18$ to 3.75.
The flux weighted, disk integrated value of gravity is 
$\log g = 3.9$, and this appears to be approximately consistent 
with the appearance of the H$\gamma$ profile (Fig.~7). 

\begin{figure} 
\begin{center} 
 {\includegraphics[angle=0,height=12cm]{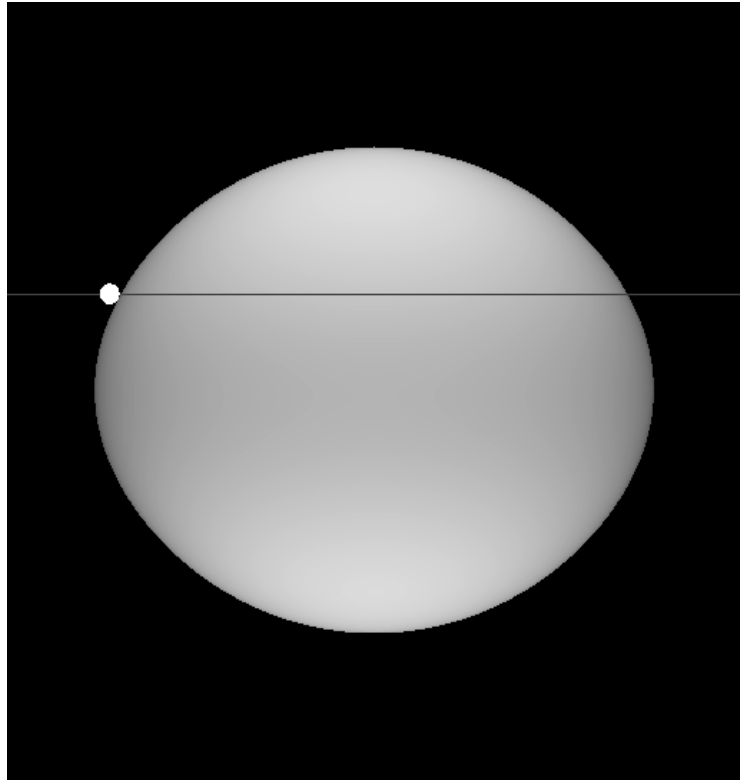}} 
\end{center} 
\caption{A model representation of the monochromatic intensity 
in the {\it Kepler} band-pass of the rotationally distorted B-star
and the small, hot companion star (shown at first contact). 
The horizontal gray line shows the derived transit path. 
\label{fig10}}
\end{figure}

\placefigure{fig10}     

We used this model image to calculate transit light curves. 
We assumed that the apparent transit path follows a trajectory 
parallel to the stellar equator with a relative velocity 
given by $2 \pi a /P$.   During phases when the disk of the 
companion is seen projected against the background B-star, 
we took the flux removed as the product of the companion area
and the specific intensity at the projected position of the 
center of companion (Mandel \& Agol 2002; Barnes 2009). For the phases 
between first and second contact, we assumed that the companion
occults a locally linear stellar limb of the B-star with a 
slope given by the Roche model shown in Figure~10, and the 
specific intensity was estimated at test positions perpendicular
to the local limb to make a numerical calculation of the occulted
flux.  This constrained model has only two free parameters:
the ratio of secondary to primary polar radius $R_2/R_{1~{\rm polar}}$ 
and the ratio $z_0/R_{1~{\rm polar}}$ of the smallest distance $z_0$ 
from the transit trajectory to the center of the B-star 
relative to the primary polar radius.
The value of $R_2/R_{1~{\rm polar}}$ sets the amount of light 
removed and hence the depth of the transit light curve, 
while the value of $z_0/R_{1~{\rm polar}}$ sets the duration 
of the transit (longer for transits near the equator with smaller $z_0$). 

We show in Figure~11 the occultation and transit light curves of KOI-81
from nine months of {\it Kepler} short cadence observations. 
These plots show the photometric fluxes normalized to unity outside 
of eclipses that we calculated by rebinning and averaging all the 
available measurements in orbital phase bins equivalent to 3~minutes duration. 
The {\it Kepler} photometry was detrended and pre-whitened for 
the primary oscillation frequency $f_1$ (\S5) before rebinning. 
The upper plot of the total occultation of the hot companion by the 
B-star indicates that the companion contributes a flux fraction of
$F_2 / F_1 = 0.00501 \pm 0.00006$ in the {\it Kepler} band-pass, 
and we renormalized the model for this extra flux in calculating 
the model transit curves.  The model curves were also convolved with 
a temporal box function to represent the bin size applied to the observations. 
Our best fit (shown in Fig.~10 and Fig.~11) was obtained with 
$R_2/R_{1~{\rm polar}} = 0.0426 \pm 0.0003$ and 
$z_0/R_{1~{\rm polar}} = 0.396 \pm 0.011$ 
(equivalent to an orbital inclination of $i = 88\fdg97 \pm 0\fdg04$). 
The individual transits display asymmetries that appear to be 
related to the pulsational phases at the times of transit, and
because the mean transit curve shown in Figure~11 represents data from 
only 12 transits, we do not attribute any significance to the 
trends in the residuals from the fitted transit light curve. 
We caution that the quoted uncertainties in the fitting parameters do not 
account for the range in the possible values for the rotation rate we adopted. 
Nevertheless, it is interesting to note that the derived 
radius of the companion $R_2/R_\odot = 0.0979 \pm 0.0019$ is close
to what we list in Table~3 based upon the spherical approximation,
mean density, and radius ratio from the {\it Kepler} project, 
and the primary radius given in Table~3 falls comfortably between
the polar and equatorial radii of the rotating model.  

\begin{figure} 
\begin{center} 
 {\includegraphics[angle=0,height=12cm]{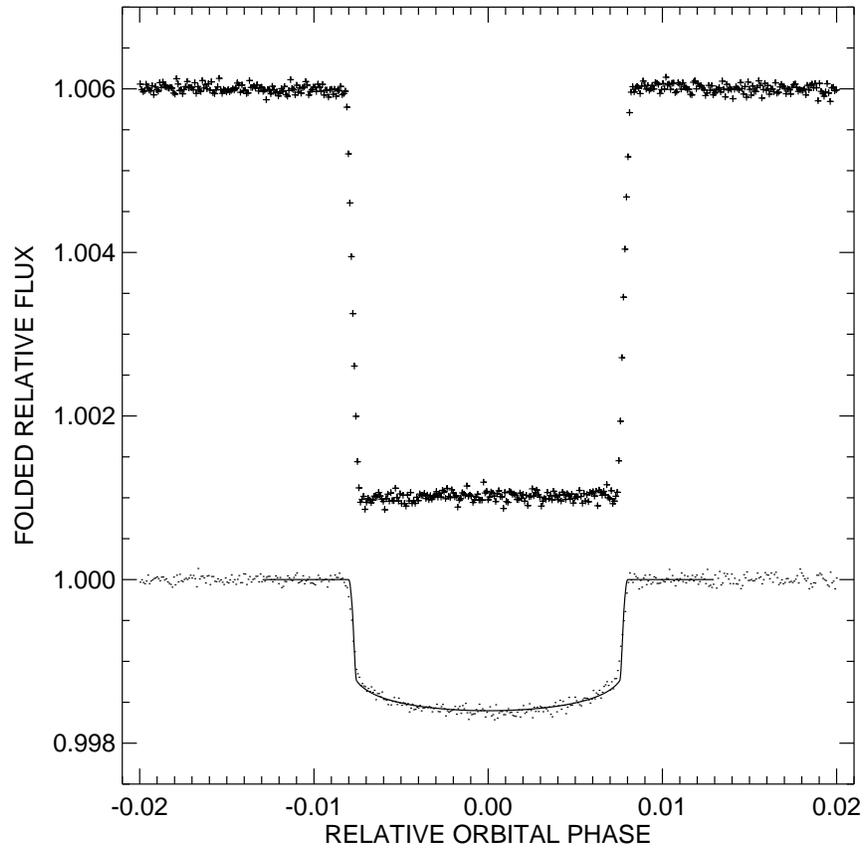}} 
\end{center} 
\caption{The phase-folded light curve of KOI-81 for the occultation 
of the companion (plus signs, shifted by half an orbit and offset
by +0.006 for clarity) and the transit of the companion (dots). 
The model transit light curve is shown by the solid line.
\label{fig11}}
\end{figure}

\placefigure{fig11}    

In \S4 we compared the observed and model spectral line depths of the 
secondary to estimate that the total flux ratio at 1370 \AA\ is 
$F_2 / F_1 = 0.05 \pm 0.02$, and using the temperatures we derived, 
we predict that this flux ratio will vary from $F_2 / F_1 = 0.14$ at 1150 \AA\ 
to $F_2 / F_1 = 0.0076$ in the {\it Kepler} optical band.  
The observed flux ratio is approximately related to radius ratio by
\begin{equation}
{F_2 \over F_1} = {f_2 \over f_1} \left({R_2 \over R_1}\right)^2
\end{equation}
where $f_2 / f_1$ is the ratio of monochromatic flux per unit area. 
According to the Kurucz flux models for our stellar parameters, 
$f_2 / f_1 = 14.5 \pm 0.5$ at 1370 \AA , so the observed flux ratio 
would then imply a radius ratio of $R_2/R_1 = 0.059 \pm 0.012$. 
Both the flux and radius ratio estimates from spectroscopy are 
somewhat larger than the corresponding estimates from the {\it Kepler} 
transit analysis, but the differences are not surprising given
the large uncertainties associated with the flux ratio from spectroscopy. 

We can also use equation 1 to infer the temperature ratio from 
the observed flux ratio in the {\it Kepler} band.  The Kurucz flux models
vary as $f \propto T_{\rm eff}^{a}$ with $a = 1.58$ for $\log g = 5$, 
$\lambda = 6430$~\AA , and $T_{\rm eff}$ in the range 10 to 25 kK. 
Thus, if we set the radius of the primary as 
$(R_{1~{\rm equator}}+R_{1~{\rm polar}})/2 = 1.11 R_{1~{\rm polar}}$, 
then we can use the values of $F_2 / F_1$ and $R_2/R_{1~{\rm polar}}$ 
from above to find a temperature ratio of $T_{2~{\rm eff}}/T_{1~{\rm eff}}=2.2$.
This is broadly consistent with the temperature estimates from spectroscopy (\S4),
$T_{2~{\rm eff}}/T_{1~{\rm eff}} = 1.7 - 2.3$.

 
\section{Discussion}                                

The {\it HST}/COS observations of KOI-81 have revealed the 
spectrum of the hot companion, the stripped down remains 
of the originally more massive star in the binary. 
Our derived mass, radius, and temperature values for KOI-81 are consistent 
with previous results.  The mass estimate of the hot, compact companion 
is slightly lower than earlier estimates ($\sim 0.3 M_\odot$; van Kerkwijk et al.\ 2010), 
solidifying its place among low mass He WDs and their immediate progenitors 
(with masses below $0.3 M_{\odot}$ where WDs have helium cores
that are too small to sustain He burning; Silvotti et al.\ 2012). 
The radius derived for the subdwarf remains essentially unchanged 
from earlier estimates, and our analysis confirms that it 
is larger than typical for He WDs (see Fig.~15 of Panei et al.\ 2007) like the 
other thermally bloated WDs found by {\it Kepler} (Rappaport et al.\ 2015).
The effective temperature range for the hot companion determined through 
spectral reconstruction is higher than the previous estimate 
of Rowe et al.\ (2010), placing it in the hot subdwarf (sdB) and WD regime.  
Of the known main sequence (MS) + sdB/He WD binaries, KOI-81 has the hottest, low mass WD 
(the others range between 9 -- 15 kK), although several of the sdB/He WD stars in 
double degenerate systems have similar effective temperatures 
(Brown et al.\ 2013; Hermes et al.\ 2014). 

The relatively long orbital period of KOI-81 suggests 
that the system widened as the result of  
conservative mass transfer following the mass ratio reversal. 
The binary probably began as a relatively close binary, 
so that mass transfer was initiated as the donor grew in size
after concluding core H burning (designated channel 1 in the 
evolutionary schemes presented by Willems \& Kolb 2004). 
Population synthesis models by Willems \& Kolb (2004) show
that the post-RLOF systems may frequently have  
a remnant mass and orbital period like those we find for KOI-81
(see their Fig.~2).  The progenitor binaries of such systems
typically have similar and low component masses and initial orbital 
periods somewhat larger than 1~d (see Fig.~2 in Willems \& Kolb 2004).
For example, van Kerkwijk et al.\ (2010) present an 
evolutionary scenario for KOI-81 that begins with a 1.8 and 
$1.3 M_\odot$ pair with an orbital period of 1.3~d. 
It is curious that the dominant pulsation period is similar to this. 
It may simply be a coincidence, but an alternative possibility is 
that this pulsation period represents the rotational period 
of the mass gainer in synchronous rotation before the onset of RLOF. 
Mass transfer from the donor would subsequently spin up the gainer, 
but it might take a long time for the angular momentum to be 
effectively redistributed into the core of the gainer star. 
We speculate that a slower rotating core may be the source of 
pulsational wave generation in the envelope of the gainer B-star. 

The mass donor remnant may have retained enough of it H-envelope to 
maintain shell H-burning longer, so that it evolves to hotter temperature before
cooling begins (Maxted et al.\ 2014; Rappaport et al. 2015).  
Based on the evolutionary scenarios presented by Silvotti et al.~(2012) 
and Rappaport et al.\ (2015), our derived mass of $0.194\pm0.020 M_\odot$ 
for the hot companion of KOI-81 is consistent with 
a low mass, pre-He WD that has not yet reached the cooling branch or 
a He core WD that is already on the cooling branch after experiencing 
episodes of shell H-burning in CNO flashes. 
However, when plotted in the ($\log T_{\rm eff}, \log g$) plane with 
the evolutionary sequences for low mass white dwarfs of Althaus et al.\ (2013),
KOI-81 appears to be in an early cooling phase of He core WD evolution. 
A specific age estimate is difficult to pin down because 
the star is located in a mass regime which experiences multiple CNO flashes 
that cause multiple loops in the ($\log T_{\rm eff}, \log g$) diagram. 
Althaus et al.\ present a grid of masses and cooling ages weighted by the 
amount of time spent in specific regions of the cooling tracks and the 
range of masses found in each region.  Using the values of $\log T_{\rm eff}$ 
and $\log g$ derived by van Kerkwijk et al.\ (2010), they find a mass of 
$0.196 \pm 0.007 M_\odot$ and cooling age of $372 \pm 186$ Myr for KOI-81 (see their Table 2). 
Our higher values of $\log T_{\rm eff}\approx 4.3$ and $\log g = 5.8$ 
correspond to a mean evolutionary track mass of $0.202 \pm 0.015 M_\odot$ and 
a cooling age $356 \pm 187$~Myr, so our derived stellar parameters are 
broadly consistent with the evolutionary model predictions.

The evolutionary path of KOI-81 is one of many that create binary systems 
with low mass WDs.  The MS + sdB binaries discovered by {\it Kepler} 
(Rappaport et al.\ 2015) and {\it WASP} (Maxted et al.\ 2014) represent
those systems with stripped-down remnants formed during the first 
RLOF phase.   These occur during the faster 
part of the pre-He WD evolutionary tracks, but it is easier 
to detect the low mass WDs at this stage because of their relatively
large luminosity (Istrate et al.\ 2014).  In fact, most of the known
low mass He WDs are those fainter WDs on the long-lived part of
the cooling tracks that are members of binaries with even fainter companions.
These are often the remanants of a second mass exchange that involves 
a common envelope stage.  They are generally short period systems that are  
sdB + WD binaries (Heber et al.\ 2003; Silvotti et al.~2012), but also include 
sdB + sdB systems (Ahmad, Jeffery \& Fullerton 2004; Kupfer et al.\ 2015),
sdB + neutron star companion (Geier et al.\ 2009; Istrate et al.\ 2014), 
and sdB + substellar companion (Geier et al.\ 2009; Geier 2015). 

The detection of the hot companion in KOI-81 provides us with 
the means to determine the stellar parameters in one example of
what must be a large population of undetected binaries 
with faint, hot companions (Willems \& Kolb 2004; Di Stefano 2011). 
The fact that the B-star component is also a pulsator opens 
up the possibility to study the stellar interior of a star 
that has been radically changed through mass transfer.  
Furthermore, an analysis of the transit shapes at different
phases in the pulsational and rotational cycles may help elucidate 
the nature of the pulsation modes and source of the rotational 
modulation (B\'{i}r\'{o} \& Nuspl 2011).  The serendipitous 
discovery of KOI-81 by {\it Kepler} has given us the opportunity 
to explore the properties of stars that have survived 
transformative mass exchange.

 
\acknowledgments 
 
We are grateful to Charles Proffitt and Denise Taylor of STScI 
for their aid in planning the observations with {\it HST}.
Support for program \#12288 was provided by NASA through a grant 
from the Space Telescope Science Institute, which is operated by 
the Association of Universities for Research in Astronomy, Inc., 
under NASA contract NAS 5-26555.
This material is also based upon work supported by the 
National Science Foundation under Grant No.~AST-1411654.
Institutional support has been provided from the GSU College 
of Arts and Sciences and the Research Program Enhancement 
fund of the Board of Regents of the University System of Georgia, 
administered through the GSU Office of the Vice President 
for Research and Economic Development.  

{\it Facilities:} \facility{HST, Kepler, Mayall, Tillinghast} 
 
 

\clearpage



\begin{deluxetable}{cccccccc}
\tabletypesize{\scriptsize} 
\tablewidth{0pc} 
\tablenum{1} 
\tablecaption{Radial Velocity Measurements\label{tab1}} 
\tablehead{ 
\colhead{Primary /}        & 
\colhead{Date}             & 
\colhead{Orbital}          & 
\colhead{$V_r$}            & 
\colhead{$V_r-\gamma_i$}   & 
\colhead{$\sigma$}         & 
\colhead{$(O-C)$}          & 
\colhead{Observation}      \\  
\colhead{Secondary}        & 
\colhead{(HJD--2,400,000)} & 
\colhead{Phase}            & 
\colhead{(km s$^{-1}$)}    & 
\colhead{(km s$^{-1}$)}    & 
\colhead{(km s$^{-1}$)}    & 
\colhead{(km s$^{-1}$)}    & 
\colhead{Source}           \\
\colhead{(1)}              & 
\colhead{(2)}              & 
\colhead{(3)}              & 
\colhead{(4)}              & 
\colhead{(5)}              & 
\colhead{(6)}              & 
\colhead{(7)}              &              
\colhead{(8)}              
} 
\startdata 
P & 55136.6338 & 0.7248 & \phn\phn\phs $   0.77$ & \phn\phs$  4.31$ & 1.57 & \phn     $ -2.34$ &TRES \\
P & 55139.6337 & 0.8504 & \phn\phn\phs $   2.03$ & \phn\phs$  5.57$ & 0.85 & \phn\phs $  0.13$ &TRES \\
P & 55143.5776 & 0.0156 & \phn\phn     $  -5.33$ & \phn    $ -1.79$ & 0.84 & \phn     $ -1.13$ &TRES \\
P & 55344.8716 & 0.4464 & \phn\phn     $  -3.46$ & \phn\phs$  0.08$ & 3.70 & \phn\phs $  2.30$ &TRES \\
P & 55347.8033 & 0.5692 & \phn\phn     $  -6.76$ & \phn    $ -3.22$ & 1.43 & \phn     $ -6.05$ &TRES \\
P & 55350.7684 & 0.6934 & \phn\phn\phs $   3.63$ & \phn\phs$  7.17$ & 1.62 & \phn\phs $  0.86$ &TRES \\
P & 55366.7730 & 0.3637 & \phn\phn     $  -8.98$ & \phn    $ -7.52$ & 1.60 & \phn     $ -2.43$ &KPNO \\
P & 55366.8595 & 0.3673 & \phn         $ -11.00$ & \phn    $ -7.46$ & 3.13 & \phn     $ -2.47$ &TRES \\
P & 55367.8714 & 0.4097 & \phn\phn     $  -3.93$ & \phn    $ -2.47$ & 1.59 & \phn\phs $  1.15$ &KPNO \\
P & 55369.7871 & 0.4899 & \phn\phn     $  -3.17$ & \phn\phs$  0.38$ & 3.69 & \phn\phs $  0.80$ &TRES \\
P & 55374.7716 & 0.6987 & \phn\phn\phs $   2.49$ & \phn\phs$  6.03$ & 3.23 & \phn     $ -0.36$ &TRES \\
P & 55375.8748 & 0.7449 & \phn\phn\phs $   6.08$ & \phn\phs$  9.62$ & 3.44 & \phn\phs $  2.88$ &TRES \\
P & 55376.7460 & 0.7814 & \phn\phs     $  10.38$ &     \phs$ 13.92$ & 2.65 & \phn\phs $  7.31$ &TRES \\
P & 55377.8023 & 0.8256 & \phn\phn\phs $   3.48$ & \phn\phs$  7.02$ & 6.23 & \phn\phs $  1.03$ &TRES \\
P & 55380.8825 & 0.9546 & \phn\phn     $  -0.89$ & \phn\phs$  2.66$ & 2.53 & \phn\phs $  0.76$ &TRES \\
P & 55401.7202 & 0.8274 & \phn\phn\phs $   5.38$ & \phn\phs$  8.92$ & 1.86 & \phn\phs $  2.97$ &TRES \\
P & 55458.7261 & 0.2150 & \phn\phn     $  -7.64$ & \phn    $ -4.10$ & 3.15 & \phn\phs $  2.47$ &TRES \\
P & 55459.6061 & 0.2518 & \phn\phn     $  -0.91$ & \phn\phs$  2.64$ & 2.08 & \phn\phs $  9.37$ &TRES \\
P & 55469.6640 & 0.6731 & \phn\phn\phs $   4.14$ & \phn\phs$  7.68$ & 3.54 & \phn\phs $  1.72$ &TRES \\
P & 55483.5960 & 0.2566 & \phn         $ -16.32$ &         $-12.78$ & 4.97 & \phn     $ -6.05$ &TRES \\
P & 55495.5829 & 0.7586 & \phn\phs     $  13.85$ &     \phs$ 17.39$ & 4.43 & \phs     $ 10.67$ &TRES \\
P & 55734.3079 & 0.7571 & \phn\phn     $  -0.33$ & \phn\phs$  8.00$ & 1.09 & \phn\phs $  1.27$ &HST/COS \\
P & 55734.4731 & 0.7640 & \phn\phn     $  -1.81$ & \phn\phs$  6.52$ & 1.09 & \phn     $ -0.19$ &HST/COS \\
P & 55775.9626 & 0.5017 & \phn\phn     $  -9.31$ & \phn    $ -0.98$ & 1.09 & \phn     $ -1.05$ &HST/COS \\
P & 55841.5269 & 0.2478 & \phn         $ -14.29$ & \phn    $ -5.96$ & 1.09 & \phn\phs $  0.78$ &HST/COS \\
P & 55841.6811 & 0.2542 & \phn         $ -15.88$ & \phn    $ -7.55$ & 1.09 & \phn     $ -0.81$ &HST/COS \\
P & 56077.7512 & 0.1415 & \phn         $ -10.35$ & \phn    $ -8.88$ & 1.84 & \phn     $ -3.65$ &KPNO \\
P & 56081.7577 & 0.3093 & \phn\phn     $  -2.55$ & \phn    $ -1.08$ & 1.87 & \phn\phs $  5.19$ &KPNO \\
P & 56082.7568 & 0.3512 & \phn\phn     $  -3.22$ & \phn    $ -1.76$ & 2.45 & \phn\phs $  3.66$ &KPNO \\
P & 56486.7101 & 0.2699 & \phn         $ -17.15$ &         $-15.68$ & 4.03 & \phn     $ -9.00$ &KPNO \\
S & 55734.3079 & 0.7571 &              $-102.81$ & \nodata          & 2.08 & \phn     $ -1.54$ &HST/COS \\
S & 55734.4731 & 0.7640 & \phn         $ -99.37$ & \nodata          & 2.12 & \phn\phs $  1.61$ &HST/COS \\
S & 55841.5269 & 0.2478 & \phs         $ 101.12$ & \nodata          & 2.42 & \phn\phs $  0.15$ &HST/COS \\
S & 55841.6811 & 0.2542 & \phs         $ 100.74$ & \nodata          & 2.81 & \phn     $ -0.21$ &HST/COS \\
\enddata 
\end{deluxetable}
  
\newpage 
 
\begin{deluxetable}{lc} 
\tablewidth{0pc} 
\tablenum{2} 
\tablecaption{Orbital Elements for KOI-81\label{tab2}} 
\tablehead{ 
\colhead{Element} & 
\colhead{Value}   } 
\startdata 
$P$~(days)                      \dotfill & 23.8760923\tablenotemark{a} \\ 
$T_t$ (HJD--2,400,000)          \dotfill & 54976.07186\tablenotemark{a} \\ 
$K_1$ (km s$^{-1}$)             \dotfill & $6.74 \pm 0.67$       \\ 
$K_2$ (km s$^{-1}$)             \dotfill & $101.18 \pm 0.73$     \\ 
$\gamma_1$[TRES] (km s$^{-1}$)  \dotfill & $-3.54 \pm 0.62$      \\ 
$\gamma_1$[KPNO] (km s$^{-1}$)  \dotfill & $-1.47 \pm 0.68$      \\ 
$\gamma_1$[COS]  (km s$^{-1}$)  \dotfill & $-8.33 \pm 0.17$      \\ 
$\gamma_2$[COS]  (km s$^{-1}$)  \dotfill & $-0.19 \pm 0.72$      \\ 
$M_2/M_1$                       \dotfill & $0.0666 \pm 0.0066$   \\ 
$a\sin i$ ($R_\odot$)           \dotfill & $50.91 \pm 0.47$      \\ 
rms$_1$ (km s$^{-1}$)           \dotfill & 2.9                   \\ 
rms$_2$ (km s$^{-1}$)           \dotfill & 1.6                   \\ 
\enddata 
\tablenotetext{a}{Fixed.}
\end{deluxetable} 

\newpage 
 
\begin{deluxetable}{lcc} 
\tablewidth{0pc} 
\tablenum{3} 
\tablecaption{Stellar Parameters for KOI-81\label{tab3}} 
\tablehead{ 
\colhead{Parameter} & 
\colhead{Primary}   & 
\colhead{Secondary} } 
\startdata 
$M/M_\odot$	                \dotfill  & $2.916 \pm 0.057$ & $0.194 \pm 0.020$   \\ 
$R/R_\odot$    \dotfill  & $2.447 \pm 0.022$\tablenotemark{a} & $0.0911 \pm 0.0025$\tablenotemark{a} \\ 
$\log g$ (cgs)                  \dotfill  & 4.13\tablenotemark{b} & 5.81\tablenotemark{b}  \\ 
$T_{\rm eff}$ (kK)              \dotfill  & $ 11.7 \pm 1.5$   & $ >19.4 \pm 2.5 $   \\ 
$V\sin i$ (km s$^{-1}$)         \dotfill  & $ 296  \pm 5$     & $ < 10 $            \\ 
\enddata 
\tablenotetext{a}{Assuming a spherical shape for the primary.}
\tablenotetext{b}{Calculated from $M/M_\odot$ and $R/R_\odot$.}
\end{deluxetable} 

\newpage 
 
\begin{deluxetable}{lccccccc} 
\tabletypesize{\small} 
\tablewidth{0pc} 
\tablenum{4} 
\tablecaption{Significant Photometric Frequencies\label{tab4}} 
\tablehead{ 
\colhead{}   & 
\colhead{Frequency ($d^{-1}$)}      &
\colhead{Amplitude ($10^{-6}$)}      &
\colhead{Phase (rad/$2\pi$)} & 
\colhead{S/N} &
\colhead{Comment} 
}
\startdata 
$f_{1}$    & $  0.722974 \pm    0.000002$ & $              257.6  \pm        4.1
  $ & $           0.822\pm         0.001$ & $              107.4$ & $$ \\
$f_{2}$    & $    1.32403 \pm    0.00001$ & $              33.4  \pm        2.8
  $ & $            0.34\pm         0.006$ & $              20.6$ & $f_{4}+f_{6}$ \\
$f_{3}$    & $    1.08445 \pm    0.00002$ & $              22.1 \pm        3.2
  $ & $            0.98\pm          0.01$ & $              20.5$ & $f_{1}+f_{6}$ \\
$f_{4}$    & $    0.96250 \pm    0.00001$ & $              34.6  \pm        3.6
  $ & $           0.22\pm          0.008$ & $              16.7$ & $$ \\
$f_{5}$    & $     2.08287\pm    0.00001$ & $              31.4  \pm        2.6
  $ & $           0.77\pm          0.006$ & $              11.4$ & $f_{rot}$ \\
$f_{6}$    & $    0.36148 \pm    0.00002$ & $              47.5  \pm        7.7
  $ & $            0.28\pm          0.01$ & $              10.6$ & $0.5f_{1}$ \\
$f_{7}$    & $    2.34753 \pm    0.00002$ & $              15.7  \pm        2.3
  $ & $            0.36\pm          0.01$ & $              8.0$ & $$ \\
$f_{8}$    & $    0.70933 \pm    0.00009$ & $              6.3  \pm        4.2
  $ & $           0.96\pm           0.04$ & $              6.9$ & $$ \\
$f_{9}$    & $    1.74647 \pm    0.00006$ & $              5.8  \pm        2.3
  $ & $           0.15\pm           0.03$ & $              6.8$ & $$ \\
$f_{10}$    & $   2.08357 \pm    0.00003$ & $              12.3  \pm        2.6
  $ & $            0.37\pm          0.05$ & $              4.5$ & $f_{rot}$ \\
$f_{11}$    & $   4.16643 \pm    0.00008$ & $              4.4  \pm        2.4
  $ & $           0.08\pm           0.04$ & $              4.3$ & $2 f_{rot}$ \\
$f_{12}$    & $     0.0837 \pm    0.0002$ & $              10.9  \pm        4.0
  $ & $           0.25\pm           0.02$ & $              3.2$ & $2f_{orb}$ \\
$f_{13}$    & $   2.08216 \pm    0.00003$ & $              10.4  \pm        2.6
  $ & $           0.09\pm           0.04$ & $              3.1$ & $ f_{rot}$ \\
\enddata 
\end{deluxetable} 

\end{document}